# Psychological Aspects of Pair Programming

A Mixed-methods Experimental Study

Marcel Valový
Department of Information Technologies
Prague University of Economics and Business
Prague, Czech Republic
marcel.valovy@vse.cz

## ABSTRACT

[Context] With the recent advent of artificially intelligent pairing partners in software engineering, it is interesting to renew the study of the psychology of pairing. Pair programming provides an attractive way of teaching software engineering to university students. Its study can also lead to a better understanding of the needs of professional software engineers in various programming roles and for the improvement of the concurrent pairing software.

[Objective] This preliminary study aimed to gain quantitative and qualitative insights into pair programming, especially students' attitudes towards its specific roles and what they require from the pairing partners. The research's goal is to use the findings to design further studies on pairing with artificial intelligence.

[Method] Using a mixed-methods and experimental approach, we distinguished the effects of the pilot, navigator, and solo roles on ($N$ = 35) students' intrinsic motivation. Four experimental sessions produced a rich data corpus in two software engineering university classrooms. It was quantitatively investigated using *the Shapiro-Wilk* normality test and one-way analysis of variance (ANOVA) to confirm the relations and significance of variations in mean intrinsic motivation in different roles. Consequently, seven semi-structured interviews were conducted with the experiment's participants. The qualitative data excerpts were subjected to the thematic analysis method in an essentialist way.

[Results] The systematic coding interview transcripts elucidated the research topic by producing seven themes for understanding the psychological aspects of pair programming and for its improvement in university classrooms. Statistical analysis of 612 self-reported intrinsic motivation inventories confirmed that students find programming in pilot-navigator roles more interesting and enjoyable than programming simultaneously.

[Conclusion] The executed experimental settings are viable for inspecting the associations between students' attitudes and the distributed cognition practice. The preliminary results illuminate the psychological aspects of the pilot-navigator roles and reveal many areas for improvement. The results also provide a strong basis for conducting further studies with the same design involving the big five personality and intrinsic motivation on using artificial intelligence in pairing and to allow comparison of those results with results of pairing with human partners.

## 1 Introduction

We believe that the psychology of pairing in software engineering should receive a new round of scientific attention due to the recent advent of artificially intelligent pairing partners in both Pilot (e.g., GitHub Copilot and ChatGPT) and Navigator (e.g., Grammarly but also ChatGPT) roles. The presented research results on pairing in software engineering can be used both to improve software engineering in university classrooms and to develop better AI pairing software.

In our study, the independent variable is represented by the chosen software engineering role, where we differentiate between the pair programming pilot and navigator roles and the solo role. Pair programming is an agile software development practice where two programmers collaborate on the same task using one computer and a single keyboard [20]. One takes on the pilot role and writes the code. The other simultaneously takes on the navigator role, thinking about the problem and conceptualizing the solution, analyzing the written code, providing feedback, and addressing issues.

The last two decades of research on pair programming have examined its effects on performance, code quality, knowledge sharing, and other aspects concluding that it is beneficial in both professional and educational settings [e.g., 3, 10, 18]. Per the seminal meta-analysis paper, pairing is faster when the task complexity is low and yields code solutions of higher quality when the task complexity is high [10]. A similar author collective also studied the moderating relation of personality on those positive effects but has not found much statistical significance [9]. Our paper differs significantly from all previous studies as it uses a revised experimental design based on [1] that distinguishes the effects in the pilot, navigator, and solo roles, not just "pair" and "solo", and invites the moderating personality factors. It also focuses on a more nuanced metric called "intrinsic motivation", which is a predictor of the previously studied variables such as performance. Finally, it studies the effects per individual, not per pair.

The dependent variable in our study is intrinsic motivation. Per the Self-determination theory, it exists in the relationship between individuals and activities [16]. Each individual is intrinsically motivated for some activities and not others. In humans, intrinsic motivation is a prototypical example of *autonomous* behavior, being willingly or volitionally done, as opposed to heteronomous actions, which are subject to external rewards or pressures (*ibid.*). Neither was allowed in our controlled experiments.

This paper will report on the psychological aspects of pair programming, or pairing, in general. Initially, we tap into the overall motivational effects of the agile practice using statistical analysis of 612 motivational inventories from four experimental sessions in two undergraduate classrooms. Afterward, we utilize the thematic analysis to discover the reoccurring themes about pairing by probing into students' feelings, attitudes, and values. The themes are pairing constellations, feedback, task complexity, soft and social skills, psychological aspects, role-specific insights, and perception of "the perfect pair programmer" and his traits.

Research questions:

*RQ1: Are both pair programming roles intrinsically more motivating to students in university classrooms than solo?*

*RQ2: What are the psychological aspects of pairing?*

## 2 Research Settings

The research was set in two advanced software engineering course classrooms taught by the author of this paper and took place over five classes, with the first being a training session. Afterward, the research continued with seven individual online interviews until the point of saturation was reached [4].

The students were subjected to four laboratory sessions of 60 minutes of net programming time each. The ($N = 35$) subjects paused every 10 minutes to self-report their motivation with a seven-item questionnaire and rotated roles in their pair. Each round, they received a new programming task but oftentimes continued the previous one because the tasks formed a continuum.

Each session lasted a total of 90 minutes, of which 30 were used for describing the tasks and completing the questionnaires.

The partners in the experimental group were designated either "pilot," who controls the keyboard and codes, or "navigator," who conceptualizes the solution and looks for defects. The participants in the control group worked on the same tasks in the "solo" role.

Each individual's motivation related to the performed role was reported after every round using a standardized seven-item questionnaire, including when programming solo in the control group. The Big Five personality test at the beginning of each session measured the subject's personality. Analysis of the personality results was not included in this preliminary study's scope, but the results were confidentially given to the subjects and indirectly queried in the qualitative interviews.

The subjects were instructed on how to pair-program during a *training session* that preceded the other four. Then, the subjects worked on predefined tasks in a static order that helped them create their individual semestral work. It was an adventure game with a graphical user interface written in Java. During the experimental sessions, students have (i) implemented the observer time for real-time updating, (ii) added a contextual menu, (iii) animated the characters in the game, and, finally, (iv) cooperated with other pairs in their six-member teams to develop the foundations of their second, more complex semestral work.

*Concerns of experimental design.*

The task difficulty was of primary concern to the experimental design's validity, which is why the first two sessions contained a "control group." In each session, the motivational levels for each 10-minute round were compared in both the control (solo programming) and the treatment group (pairing) to confirm that there was, in fact, no correlation between the specific tasks and self-reported intrinsic motivation. This is consistent with [18] and our qualitative results.

The second concern, the subjects were of various abilities. If performances were measured, the subjects' abilities could have diminished the measured effect, like in [3], as *performance = ability x motivation* [12]. We avoided that by measuring intrinsic motivation instead, which does not directly relate to ability unless the task is outside the individual's perceived competency levels [16].

Lastly, the uncertain personal compatibility of pairs posed a threat. Pairs were allocated pseudo-randomly and irrespective of personality, like in two previous experimental studies, one with 564 students, where 90 % of pairs reported compatibility [11], and another with 1350 students, and 93 % reported compatibility [19].

## 3 Methods

### 3.1 Instruments

Recent studies, such as [6, 7], advocate the use of instruments coming from psychology and related fields for systematic studies in the IT discipline and emergent "behavioral software engineering" field [13].

Intrinsic motivation related to the programming role was measured using a self-report questionnaire Intrinsic Motivation Inventory at the end of each 10-minute-long round of which there were six in each session. The questionnaire was developed and first used by Ryan in a series of psychological experiments in the 1980s [15]. Its first subscale that was used consists of seven Likert-scale items, which are factor-analytically coherent and stable across various tasks, conditions, and settings, and assesses participants' subjective experience related to target activity in laboratory experiments [17].

The final questionnaire deployed in the experiments also contained the Big Five personality measurements. All instruments were implemented in their original English version.

### 3.2 Quantitative Analysis

We used scripts written in R language v4.2.2 and open-source libraries for statistical tests and computations.

Descriptive statistics were used to describe the sample and provide a demographic overview. Subsequently, *the Shapiro-Wilk test* evaluated the reported intrinsic motivation values in various programming roles for their normality. Assuming the normal distributions, a one-way parametric test of variance (ANOVA) was used to verify the statistical significance of the derived relations between the chosen programming role and reported intrinsic motivation.

### 3.3 Thematic Analysis

The inquiry framework presented in this article uses thematic analysis developed for use within a qualitative paradigm, subjecting data to analysis for commonly recurring themes. It is theoretically flexible, and the author has chosen the inductive (bottom-up) way of identifying patterns in the data instead of the deductive (top-down) approach. Inductive analysis was used to code the data without trying to fit it into a pre-existing coding frame or the researcher's analytic preconceptions.

Seven steps by [5] were applied flexibly to fit the research questions and data: *transcribing, becoming familiar with the data, generating initial codes, discovering themes, reviewing themes, defining themes,* and *writing up.* Transcribing (step 1) can be seen as a critical phase of interpretative data analysis, as the meanings are created during this thorough act [4]. A professional tool (Descript, v52.1.1) was used for initial pre-processing to generate basic transcription from voice recordings and to remove filler words like "um." Final transcripts were imported into a computer-aided qualitative data analysis tool (MAXQDA, v22.3.0), coded, and analyzed for themes (steps 2-6).

## 4 Results

### 4.1 Quantitative Results

Five of the 35 students in two university classrooms were females, and thirty were males. The students' software engineering experience ranged from zero to five years, with a mean of 1.6 years and a standard deviation of 1.3 years.

We can assume the normality of the reported motivational levels from the output obtained in *the Shapiro-Wilk* normality test ($W$ = 0.98694, $p$-value = 0.4713). The $p$-value is greater than 0.05. Hence, the distribution of the reported intrinsic motivational levels is *not significantly different* from the normal distribution and permits the usage of parametric tests such as ANOVA.

From the output obtained in a one-way ANOVA test ($F$-value = 6.618, $p$-value = 0.00206 **), where the $p$-value is less than $\alpha$ = .05, we successfully reject the null hypothesis that the groups representing chosen programming roles are similar in reported motivational level. This translates to the variance between those groups being large relative to their internal variance, and, therefore, concluding that this nominal variable (chosen programming role) holds distinct relations toward motivation.

The reported intrinsic motivational levels in the programming roles of Pilot ($\mu$ = 28.41, $\sigma$ = 2.91), Navigator ($\mu$ = 27.87, $\sigma$ = 3.24), and Solo ($\mu$ = 25.11, $\sigma$ = 2.66) are reported in Table I with their respective means and standard deviations.

TABLE I.    MOTIVATION IN PILOT, NAVIGATOR, AND SOLO ROLES

|   | Pilot | Navigator | Solo |
|---|---|---|---|
| μ | 28.41 | 27.87 | 25.11 |
| σ | 2.91 | 3.24 | 2.66 |

### 4.2 Qualitative Results

Seven overarching themes were discovered that depict the psychological aspects of pair programming in educational settings. The themes and the codes that established them are displayed in Table II. The first column identifies the themes. Authentic excerpts in the second column capture the students' mindsets. The codification elements that led to the creation of the themes are listed in the third column. Finally, observed subthemes are presented in column number four.

## 5 DISCUSSION

The main aim of our preliminary study was to shed light on the psychological aspects of pairing in software engineering and to find ways of improving its utilization in educational settings while building a knowledge base for AI pairing software research. Prior experimental designs on the topic were extended with novel approaches, incorporating the Self-determination framework [16] and distinguishing all effects per individual, as opposed to by pair.

### 5.1 Answering the research questions

The following discussion will provide answers to the research questions and recapitulate the observed themes, codes, and subthemes. For more insight, the reader should refer to Table II.

*5.1.1 RQ1: Are both pair programming roles intrinsically more motivating to students in university classrooms than solo?* By quantitative results, the pilot-navigator roles are more intrinsically motivating ($\mu$ = 28.41 and 27.87, respectively) than both members coding simultaneously ($\mu$ = 25.11), which is a vital sign and

TABLE II. SEVEN THEMES FORMING THE PSYCHOLOGICAL ASPECTS OF PAIR PROGRAMMING (PP)

| Theme | Participants' Quotes | Codes | Subthemes |
|---|---|---|---|
| Pairing Constellations | "Pairing with a familiar person allows you to anticipate their reactions and, therefore, be more honest with them, which is more comfortable." (P1)<br>"The best part was when I could establish a good rapport with my partner." (P1)<br>"If there were a system by which the partners were assigned to each other and stayed like that the whole semester, it would be nicer. The system would be difficult to design. Maybe something like speed-dating where you could match personal compatibilities." (P1)<br>"The more experienced should give thoughts to pilot" (P6) | "Good pairing constellations create friendships"<br>"If socially compatible, you can pair for longer"<br>"Code style differences yield pros (learning) and cons (conflict)"<br>"Contrast in extraversion can be fatal"<br>"If one is low in agreeableness and the other high in neuroticism, they can close down"<br>"Problematic combinations lie in the extremes"<br>"Maybe it is a bit of an alchemy"<br>"I perform better with someone I know"<br>"Same pace of work is vital"<br>"Give everyone a partner with a similar skill level" | Skill-level Compatibility<br>Personal Partner Familiarity<br>Extremes<br>Contrasts<br>Friendships<br>Persistence |
| Feedback | "Feedback is important for everyone to progress in anything; whether negative or positive, it must be told." (P4)<br>"Sometimes he was really dominant, so I had to tell him 'Yes, I can do this on my own.'" (P6)<br>"Of course, communication is an integral part of PP. During the conversation, you can discover errors, discuss the strategy, find a solution on which both agree, and share knowledge in a great way." (P2) | "A good feeling of connection"<br>"Helping makes you feel better"<br>"Training for becoming a manager"<br>"Therapeutic effects – helps with mood"<br>"The instructor should show correct solution after each round"<br>"Explain what he was doing to value me"<br>"Partner's opinion about my work and me is important"<br>"Every person provides a differing and valuable feedback" | Helping<br>Connection<br>Training<br>Therapy<br>Valuing each other<br>Feedback is progress |
| Soft & Social Skills | "Programming consists of two skills: programming something and asking questions perfectly. The latter is really difficult." (P1)<br>"PP was even better than on Discord or chat. You understand body gestures, eye signals, and so on." (P5) | "It was motivating we did it together"<br>"Talking about a problem is difficult / makes me learn new skills / fortifies knowledge / requires simplifying the solution"<br>"A small simulation of real-world settings"<br>"Importance of discovering how and whom to work with" | Most difficult<br>Together<br>Real-world training |
| Psychological Aspects | "Almost everything can be solved in pairs, from programming to your emotional state." (P2)<br>"It was interesting to understand what happens in other people's brains. How they see and solve the issue." (P3)<br>"Very interesting to compare what you would plan to do with what your partner is suggesting." (P7) | "Being part of an experiment"<br>"Interesting to follow myself, how I feel"<br>"Roles and being observed push you out of your comfort zone"<br>"Pairing is 1000 % more fun than solo"<br>"Experimenting leads to discovery of new things"<br>"Personality traits determine the role preference" | Everything is easier<br>Feelings<br>Mental processes<br>Traits |
| Role-specific Insights | "It was cool that I could only focus on writing the syntax and let the Navigator come out with the logical parts." (P1)<br>"I liked that I could do my own research without someone watching in the solo role." (P1)<br>"Both roles check for a different type of errors." (P2)<br>"I like analyzing things and when I did not have to worry about coding, I had so much space in my brain, I had a different view and saw probably the best approach." (P4)<br>"It was uncomfortable when she was navigating. I like being in control of the situation. Just turned my mind off and did the things I was told to." (P5)<br>"As the navigator, I felt like a leader. The whole situation was in my hands. It motivated me a lot!" (P5) | "Pilot depended on me – motivating"<br>"Problem: introverts taking on the navigator role and not talking much"<br>"The navigator has the final say"<br>"Partner tries to implement your own thoughts"<br>"Responsibility for outcome is on the navigator"<br>"I felt I was the main point of responsibility"<br>"I looked at it not as a coder but as the analytic and reviewer"<br>"Found ways I wouldn't have found in the pilot role"<br>"Pilot should be open when someone presents him new ideas"<br>"Navigator does not necessarily have to devise novel ways to reach the destination"<br>"In pilot, you lack space for creativity or self-expression"<br>"When navigator, you feel you control the situation" | Trade-offs<br>Responsibility<br>Control<br>Fresh view<br>More energy<br>Brain space<br>Analytics<br>Suffering |
| Task Complexity & Time Constraint | "I think the time limit was good because when not sitting at the computer for long, I am losing focus." (P1)<br>"Not having the same amount of seat time could spark a sense of unfairness or jealousy." (P1)<br>"In my case, I was three times more productive in pairing than when doing the same task at home." (P3)<br>"Timer is confusing because you leave your job unfinished, and now you have to take on another role." (P5)<br>"We were short on time, struggling with others, and after a while, we caught up with the class. That was exciting!" (P6) | "Difficulty does not play a key role."<br>"Task difficulty does not influence the *suitability* of PP"<br>"Task type does not influence the *suitability* of PP"<br>"Time limit should be in every PP session"<br>"Limits help you contain yourself within and plan efficiently"<br>"Scratch the timers; I need time to think"<br>"Prefer to rotate after the task is finished"<br>"Especially tasks of profound impact should be solved in pairs"<br>"Give extra time"<br>"In pairs, all tasks seem easier" | Difficulty<br>Task Type<br>Containment<br>Productivity<br>Efficiency<br>Stimulus<br>Fairness<br>Stress<br>Finishing |
| Perfect Pair Programmer's Traits | "Be communicative, have quick learning ability, tendency to display self-discipline, be calm, willing to compromise their interest with others, and emotionally stable." (P2) | "Understanding", "Quick-witted", "Tolerant to mistakes"<br>"Good social skills", "Easy-going"<br>"Humble!", "Friendly and make you feel good working with"<br>"High extraversion", "High agreeableness", "High openness", "Low neuroticism", | |

implies that universities should include pair programming in software engineering courses. When people are autonomously motivated, their interests and values align with their actions, which become *biologically* distinct from controlled behaviors. Consequently, autonomous motivation leads to higher creativity, better problem-solving (*e.g.,* thinking outside the box), increased performance (particularly in heuristic activities like programming), positive emotions, and psychological and physical wellness [16]. Noteworthy, some prior research has shown that there are exceptions, and some students prefer pairing in only one of the roles over solo and would rather be solo than in the other pairing role [2]. Our quantitative data, however, did not show that.

*5.1.2 RQ2: What are the psychological aspects of pair programming?*

The discussion about the answers to the second research question will be divided into seven parts by its areas: pairing constellations, feedback, soft and social skills practice, psychological aspects, role-specific insights, task complexity and time constraint, and discovering the "perfect programmer's" traits.

*5.1.2.1. Pairing Constellations.* The richest of the identified themes is pairing constellations with its three subthemes: skill level, personal compatibilities, and partner familiarity, which all significantly influence motivation and productivity in pairs.

First, social compatibility is vital and allows pairing longer. Great pairing constellations are characterized as those where the partners' skill level and pace of work are the same or similar. Such pairing constellations lead to new friendships. On the other hand, if partners are familiar upfront, it opens ways to communicate more directly and effectively, thereby producing better results.

Concerns have been raised about forming detrimental pairs where members score extreme values on specific big five dimensions, such as pairing someone with high sensitivity to negative emotion (neuroticism) with someone low on agreeableness. Similar extraversion and also skill levels are vital.

A speed-dating-like system was proposed as a solution to help form the pairs at the beginning of the semester and solve the "bit of an alchemy" problem of pairing constellations. The students could test their character compatibilities in brief talk sessions, and the instructor would assign the pairs on resulting preference cards.

*5.1.2.2. Feedback.* The second theme is about feedback, which is suggested by [14] to play an essential role in software engineering, especially in motivation, and our experimental participants also stressed it. They listed a variety of positive effects stemming from providing feedback during pairings, such as the valuation of the other, the facilitation of their improvement, connective therapy, or enablement of progress. On another note, participant P6 expressed that his dominant partner had to be suppressed: "Yes, I can do this on my own."

*5.1.2.3. Soft and Social Skills.* The training of soft and social skills was found to be of crucial importance and a valued benefit yet hard to master, as participant P1 voiced: "Programming consists of two skills: programming something and asking questions perfectly. The latter is really difficult." Participants felt they were simulating real-world settings in classrooms.

*5.1.2.4. Psychological Aspects.* Participants felt more empowered in pairs, stating that "Almost everything can be solved in pairs, from programming to your emotional state." (P2)

They also enjoyed the experimental psychological introspections, concluding that the experimental pair programming was emotional and fun and that they believe personality traits profoundly affect role preferences. For example, four interview participants' voices were marked by the code "roles and being observed push you out of your comfort zone." They expressed that in such cases, they performed better, which is commonly known as "the Hawthorne effect."

Also, participants P3 and P7 found it very interesting to be able to compare their own thoughts with the thoughts of someone else, "Very interesting to compare what you would plan to do with what your partner is suggesting." (P7). Such insights are transferrable to the AI pairing software, suggesting that the user experience of comparing is essential and should receive particular attention.

*5.1.2.5. Role-specific insights.* Participants differed in their views on the benefits and negatives of the three distinct programming roles. For example, some enjoyed the "free space in their head" that pairing provided by splitting the activities per person in half, while others suffered from "being confined" in just a subset of the activities of software engineering, which is inherently multitasking in its nature. Contrarily, they all praised attaining responsibilities, feeling of control, and perceived importance.

Of particular interest is that splitting roles allows for developing novel, otherwise inaccessible solutions, as the code "found ways I wouldn't have found in the pilot role" explains. Also, a command hierarchy originates in role splitting, where "the navigator has the final say."

Lastly, it is essential to say that the participants found that people of some personality traits are fitter for one role and not the other, e.g., code "problem: introverts taking on the navigator role and not talking much."

*5.1.2.6. Task Complexity and Time Constraint.* The participants voiced opposing opinions on the strict time limit for staying in their roles. While it provided a fair share of "seat time," helped with efficiency, and prevented attention deficiencies, it also "confused" those who had not finished their tasks within the limit. Some proposed to make the time limit longer, e.g., 15 minutes per task, to allow for some research time if the pair get lost in the task. Proposed improvements also included revealing the right solution at the end of each round so that everyone could proceed together. Task type and difficulty were found not to affect pair programming.

*5.1.2.7.* Last theme was about discovering the traits of the perfect programmer and, in effect, finding ways to be a better pairing partner or construct or parametrize a better AI pairing software in ways that it "behaves" toward the user. The perfect pairing partner was described as someone humble, tolerant of mistakes, friendly, makes others feel good, and has the tendency to show self-discipline. In the Big five terms, he would score high on extraversion, agreeableness, and openness but low on the neuroticism scale. This is in accord with the results of a previous study [2], where members of the personality cluster scoring high on extraversion and agreeableness were found to be great navigators, and members of the cluster characterized by high openness strongly preferred to be in the pilot role.

### 5.2 Threats to validity

The subjective nature of interpretation poses threats to the validity of our qualitative results. However, the data have been processed systematically and in an epistemological way that introduces as little subjective bias as possible.

Internal validity in our research context mainly refers to the suitability of our data sets for the application of the selected statistical methods. This discussion was covered in the methods.

### 5.3 Limitations and generalizability

More research is needed to fully understand the psychological aspects of pairing and pair programming to understand how members of different personality clusters pair with different AI software representing the roles of pilot or navigator. However, some results, such as the "perfect programmer's" big five dimensions, can already be reused in the development of AI pairing software. Unfortunately, the results of studies on pair programming in university classrooms may not necessarily generalize to professional software development settings, where different setup is more commonly used, such as switching roles deliberately or after finishing a task and also pairing with someone else after a switch of the role.

## 6 CONCLUSION

This preliminary study mapped the psychological aspects of pair programming, analyzed ways of increasing students' intrinsic motivation and improving the pair programming practice in educational settings, and tapped into the psychology of pairing.

Pilot-navigator roles are more motivating than both members coding simultaneously. The areas for improvement of pairing include task complexity, pairing constellations, feedback, soft and social skills practice, psychological aspects, role-specific insights, and following the "perfect programmer's" traits.

With further advancement of theoretical foundations around the psychology of pairing in software engineering and the development of artificially intelligent pairing programs, the further research should orient toward the experimental study of the cooperation between humans and artificial intelligence.

### ACKNOWLEDGMENTS

This work was supported by an internal grant funding scheme (F4/61/2023) administered by the Prague University of Economics and Business.

### REFERENCES

[1]  Valový, Marcel. "Experimental Pair Programming: A Study Design and Preliminary Results." In *Proceedings of the 33rd Psychology of Programming Interest Group (PPIG'22)*, 2022: 107-112
[2]  Valový, Marcel. "Effects of Pilot, Navigator, and Solo Programming Roles on Motivation." *New Perspectives in Software Engineering: Proceedings of the 11th International Conference on Software Process Improvement (CIMPS 2022)*. Cham: Springer International Publishing, 2022: 84-98.
[3]  Arisholm, Erik, Hans Gallis, Tore Dybå, and Dag IK Sjøberg. "Evaluating pair programming with respect to system complexity and programmer expertise." *IEEE Transactions on Software Engineering 33.2 (2007)*: 65-86.
[4]  Bird, Cindy M. "How I stopped dreading and learned to love transcription." *Qualitative inquiry 11, no. 2 (2005)*: 226-248.
[5]  Braun, Virginia, and Victoria Clarke. "Using thematic analysis in psychology." *Qualitative research in psychology 3, no. 2 (2006):* 77-101.
[6]  Feldt, Robert, Lefteris Angelis, Richard Torkar, and Maria Samuelsson. "Links between the personalities, views and attitudes of software engineers." *Information and Software Technology 52, no. 6 (2010)*: 611-624.
[7]  Graziotin, Daniel, Per Lenberg, Robert Feldt, and Stefan Wagner. "Psychometrics in behavioral software engineering: A methodological introduction with guidelines." *ACM Transactions on Software Engineering and Methodology (TOSEM) 31, no. 1 (2021)*: 1-36.
[8]  Guest, Greg, Arwen Bunce, and Laura Johnson. "How many interviews are enough? An experiment with data saturation and variability." *Field methods 18, no. 1 (2006)*: 59-82.
[9]  Hannay, Jo E., Erik Arisholm, Harald Engvik, and Dag IK Sjøberg. "Effects of personality on pair programming." *IEEE Transactions on Software Engineering 36, no. 1 (2009)*: 61-80.
[10] Hannay, Jo E., Tore Dybå, Erik Arisholm, and Dag IK Sjøberg. "The effectiveness of pair programming: A meta-analysis." *Information and software technology 51.7 (2009)*: 1110-1122.
[11] Katira, Neha, Laurie Williams, Eric Wiebe, Carol Miller, Suzanne Balik, Ed Gehringer. "On understanding compatibility of student pair programmers." *Proceedings of the 35th SIGCSE technical symposium on Computer science education.* 2004.
[12] Latham, Gary P. *Work motivation: History, theory, research, and practice*. Sage, 2012.
[13] Lenberg, Per, Robert Feldt, and Lars Göran Wallgren. "Behavioral software engineering: A definition and systematic literature review." *Journal of Systems and software 107 (2015)*: 15-37.
[14] Sach, Rien. *The Impact of Feedback on the Motivation of Software Engineers*. Open University (United Kingdom), 2013.
[15] Ryan, Richard M. "Control and information in the intrapersonal sphere: An extension of cognitive evaluation theory." *Journal of personality and social psychology 43.3 (1982)*: 450.
[16] Ryan, Richard M., and Edward L. Deci. "Self-determination theory." *Basic psychological needs in motivation, development, and wellness (2017)*.
[17] Self-Determination Theory. 2022. *Intrinsic Motivation Inventory*. (January 2023). Retrieved January 12, 2023 from http://selfdeterminationtheory.org/intrinsic-motivation-inventory/
[18] Vanhanen, Jari, and Casper Lassenius. "Effects of pair programming at the development team level: an experiment." In *International Symposium on Empirical Software Engineering, (2005).*, pp. 10-pp. IEEE, 2005.
[19] Williams, Laurie, and Robert R. Kessler. *Pair programming illuminated*. Addison-Wesley Professional, 2003.
[20] Williams, Laurie, Lucas Layman, Jason Osborne, Neha Katira. "Examining the compatibility of student pair programmers." *AGILE 2006 (AGILE'06)*. IEEE, 2006